\documentclass[11pt]{article}
\usepackage[margin=1in]{geometry}
\usepackage[T1]{fontenc}
\usepackage[utf8]{inputenc}
\usepackage{lmodern}
\usepackage{setspace}
\usepackage{hyperref}
\usepackage{url}
\usepackage{graphicx}
\usepackage{booktabs}
\usepackage{algorithm}
\usepackage{algpseudocode}
\usepackage{enumitem}
\usepackage{csquotes}
\usepackage{amsmath}

\hypersetup{colorlinks=true,linkcolor=blue,citecolor=blue,urlcolor=blue}
\title{The Ephemeral Web and the Case for Proactive Archiving}
\author{Meliksah Yorulmazlar \\
Department of Computer Science,\\
Rensselaer Polytechnic Institute \\
 Troy, NY 12180, USA \\
\texttt{yorulk@rpi.edu}}
\date{17 May 2026}

\begin{document}
\maketitle

\begin{abstract}
The web is often treated as a durable record of institutional and social life, yet in practice it is fragile, revisable, and frequently ephemeral. Domains change, redesigns erase earlier material, institutions relocate, maintainers graduate, platforms impose silent limits, and periods of political instability can interrupt digital access entirely. This paper argues that archiving should not remain a niche activity practiced by a few specialists at the margins, but should become a proactive part of website maintenance. I motivate this claim through a case study centered on the Pakistan Embassy International School and College Tehran, whose domain, visual identity, leadership, and physical location all changed within a short period after my graduation. In response, I built and deployed a lightweight automated archival system using Python and GitHub Actions to submit pages and media from the site to the Internet Archive's Wayback Machine. The project shows both that archival preservation can be automated with modest infrastructure and that archival systems are themselves vulnerable to interruption, as illustrated by GitHub's automatic disabling of scheduled workflows after repository inactivity. Drawing on personal experience with internet shutdowns in Iran, open-source sustainability lessons from RPI's RCOS, and the operational history of the archiver, I argue that the ephemerality of the web is not an exception but a structural condition. If digital societies wish to preserve institutional memory and public history without leaving preservation to chance, proactive archiving should become a commonplace part of website maintenance.
\end{abstract}

\section{Introduction}
The internet is often imagined as permanent. Once something appears online, people assume it will remain retrievable indefinitely. Yet lived experience suggests otherwise. URLs change, websites are redesigned, older content disappears, institutions migrate to new platforms, and digital traces that once seemed secure are quietly lost. Large-scale measurement bears this out: Pew Research Center found that 38\% of webpages that existed in 2013 were no longer accessible by October 2023, and that link decay affects not only the broader web but also government, news, and reference ecosystems \cite{pew-linkrot}.

My own confrontation with this problem came through the Pakistan Embassy International School and College Tehran (hereafter, \emph{the school}). Within roughly two years after completing my A-Level education there, the school underwent sweeping changes. Its website moved from one domain to another, the older site's visible contents were lost, the principal changed, and the institution moved into a different building in Pasdaran, Tehran. These changes were significant not only administratively, but historically. They altered the visible record of what the institution had been.

This mattered to me personally for a second reason as well: after graduating, the school website became one of my main ways of remaining in touch with what was happening there. It served as a tether to the institution after I had left it. When I joined the school in 2018, I witnessed it transition into a worse building. Later, after I had graduated, I was genuinely happy to see that it had finally moved into a more appropriate space for students. But that transition also sharpened a realization: institutions often preserve their present while neglecting their past. During website redesigns, domain changes, and broader transitions, they may inadvertently remove institutional memory without intending to do so.

In December 2025, during a phone call with my best friend Altay, I told him that I wanted to automate the saving of the school's photos and pages so that future generations would not face the same loss of memory that our generation had faced. When we saw that the school had finally moved into a good building, we agreed that the memory of the school should be preserved for those who would come after us.

What began as a personal preservation effort quickly expanded into a broader reflection. The problem was not merely that one school website had changed. The deeper issue was that the web itself is ephemeral, and that society still treats archiving as exceptional rather than integrating it into ordinary website maintenance. This paper develops that claim through a concrete case study: the design, deployment, and lessons of an automated archiver built to preserve a school website under conditions of institutional transition, infrastructure fragility, and political uncertainty.

\section{Background and Motivation}
\subsection{Website redesign as quiet historical erasure}
Website redesign is usually framed as improvement. New branding, updated pages, a cleaner layout, and a modern domain are treated as progress. Yet redesign can also function as quiet historical erasure. In the physical world, institutional change usually leaves visible traces: an old building may still stand, a sign may remain, or records may survive elsewhere. On the web, by contrast, earlier forms can disappear almost completely. A redirect from an old URL to a new one preserves reachability while erasing chronology. The visitor still arrives, but the earlier form of the institution may vanish so thoroughly that its loss is not even apparent unless someone consults an archive.

In the case that motivated this work, the school's older domain, \url{https://www.pisctehran.com}, was replaced by \url{https://peisctehran.com}. The old URL now redirects to the new one, but the prior website's contents were not preserved in a visible way. In effect, continuity of access concealed discontinuity of memory.

\subsection{Living in Iran and experiencing digital fragility}

My concern about preservation was shaped by living in Iran from 2018 to 2023 and by later events there. During my time in Iran, I experienced two major internet shutdowns: one in November 2019 and another during the 2022--23 Mahsa Amini protests. In 2026, nationwide disruption and blackout conditions again became severe; reporting from the UK government described a near-total communications blackout from 8 January 2026 onward, while NetBlocks documented prolonged and repeated nationwide disruption, including a renewed near-total collapse after the late-February military escalation \cite{uk-iran-bulletin,netblocks-iran}. These events made it impossible for me to regard digital access as guaranteed.

The school website also appeared to rely on WordPress-related infrastructure. Under prolonged filtering and instability, this raised a practical concern: even if a website remained publicly reachable, dependence on externally hosted services could make updates much harder for its owners. In other words, a site may stay online while becoming difficult to maintain. I present this point as an informed operational observation rather than as a controlled network measurement study. The broader lesson is that digital fragility is not only about disappearance; it is also about the weakening of the conditions needed for continued maintenance, revision, and preservation.

\subsection{Inspiration from RCOS, Concerto, and QuACS}
The design of the project was also influenced by experiences at Rensselaer Polytechnic Institute, particularly through RCOS, the Rensselaer Center for Open Source. One recurring lesson in RCOS is that many student-built systems face an Achilles' heel: once the students graduate, the project often loses maintainers and risks going extinct.

I saw this dynamic vividly while helping the director of RCOS, Wesley Turner carry televisions from Lally Hall to Amos Eaton so that the Concerto project could be revived. Concerto is an open-source web-based digital signage system used to display events, weather, time, and other information on campus screens \cite{concerto-overview}. The experience left a strong impression on me because it made tangible a simple truth about student projects: they often become vulnerable after graduation, not because they lack value, but because the people who sustained them are no longer there to maintain them.

I also saw a more durable model in QuACS, the course scheduling site used by RPI students. In an interview with \emph{The Polytechnic}, its developers explained that they did not want to maintain a server after graduation and therefore designed the site to run hands-off via GitHub Actions, precisely so that it could survive them \cite{quacs-poly}. That design philosophy strongly influenced my own thinking. If preservation infrastructure can be made hands-off, it has a better chance of surviving its creators.

\section{System Design}
In mid-January 2026, I implemented an automated archiver in Python and deployed it through GitHub Actions in a public repository \cite{archiver-repo}. The system was designed to discover internal links on the school website and submit them to the Internet Archive's Wayback Machine.

\subsection{Core logic}
The archiver begins from a seed domain and recursively discovers additional internal links. It filters out CSS and JavaScript resources while including media-like resources such as files in \texttt{wp-content}, since these often contain photographs and PDFs with historical value. After collecting the set of discovered links, it shuffles them before archival submission.

This randomization is important because the workflow runs for only about 3 hours and 55 minutes, far less than the time required to archive the entire discovered site in a single pass. In practice, archiving the full set of links would require closer to 10--12 hours, while GitHub Actions imposes a six-hour maximum per run. If the links were always processed in the same order, the archiver would repeatedly favor the earliest portion of the site while later sections would receive much less attention. By shuffling the discovered links before each run, the system gives different parts of the site a roughly equal chance of being archived over repeated runs.

\begin{algorithm}[h]
\begin{algorithmic}[1]
\caption{Link Discovery and Archival Submission Workflow}
\State Set runtime cutoff $T \gets 3\text{h }55\text{m}$
\State Initialize $Q$ with the seed URL and initialize an empty set $F$ of discovered links
\While{$Q$ is not empty}
    \State Pop the next URL $u$ from $Q$
    \If{$u \notin F$}
        \State Fetch $u$ and extract quoted internal links from the page source
        \ForAll{internal links $v$ discovered on $u$}
            \If{$v$ is a media/resource path (e.g., in \texttt{wp-content})}
                \State Add $v$ directly to $F$
            \ElsIf{$v$ is not CSS/JS and $v \notin F$}
                \State Add $v$ to $Q$
            \EndIf
        \EndFor
        \State Add $u$ to $F$
    \EndIf
\EndWhile
\State Shuffle the elements of $F$
\ForAll{$u$ in shuffled $F$}
    \State Try to submit $u$ to the Wayback Machine save endpoint
    \If{a connection error occurs}
        \State report the error
    \EndIf
    \If{elapsed runtime exceeds $T$}
    \State break
\EndIf
\EndFor
\end{algorithmic}
\end{algorithm}

\subsection{Runtime and deployment model}
The project was deployed on GitHub Actions with the workflow scheduled to run six times per day. Each run was capped at approximately 3 hours and 55 minutes, deliberately below the single-run ceiling the project was designed around. The intent was to sustain continuous archival activity without appearing abusive to the platform.

The repository was public, and the system relied on the Wayback Machine's save endpoint to request archival captures. In this sense, the project was symbiotic rather than self-contained: it used GitHub as automation infrastructure and the Internet Archive as preservation infrastructure.

\subsection{Feasibility}
From mid-January to mid-April 2026, the system ran approximately 580 times, averaging roughly four hours per run and consuming nearly 2{,}500 hours of execution time in aggregate. These numbers are important not because they are large in an absolute sense, but because they demonstrate that proactive archiving is already feasible using widely available, low-cost infrastructure.

The project did not require a custom server, paid cloud orchestration, or institutional backing. A single individual using public tools was able to create a persistent preservation process.

\section{A Case Study in Institutional Memory}
The preserved site was not chosen at random. It represented a concrete case of institutional transition.

The school changed domains. It changed principals. It moved buildings. The combination of these changes created a risk that the school's visible past would become flattened into its present. The project emerged from the belief that future generations of students should be able to see how the institution looked before these transitions.

This is what makes the case broader than one school. Across the web, schools, departments, labs, nonprofits, newspapers, and local institutions frequently redesign pages without preserving earlier versions. What disappears is not just code, but context: photographs, design choices, evidence of place, and traces of what an institution once was.

The value of archiving in this case was therefore not just recovery after loss, but preservation before neglect. Even where some images remained online, the project treated them as historically vulnerable rather than safely permanent.

\section{Operational Lessons}
\subsection{Archiving systems are also ephemeral}
One of the most important lessons of the project was that preservation infrastructure itself is fragile.

The system was designed to be hands-off, yet GitHub automatically disables scheduled workflows in public repositories after 60 days without repository activity \cite{github-disable}. I only discovered this 12 days later after it was disabled. In other words, the archiver did not fail because the website vanished or because the code was broken. It failed because the automation platform silently imposed a lifecycle rule.

This revealed a deeper principle: archiving requires continuity, and continuity itself must be monitored. An archival system should not only preserve target pages; it should also make it easy to tell whether the preservation process itself is still functioning. Warnings, heartbeat checks, and interruption alerts are therefore essential components of serious preservation infrastructure.

\subsection{Website maintainability under instability}

Another lesson concerns the difference between a website being visible and a website being maintainable. Internet shutdowns, filtering, and dependence on foreign services can leave a site publicly reachable while making it much harder for its owners to update, revise, or preserve. The web's ephemerality is therefore not limited to dead links. It also includes contingent liveness: pages remain online, but the pathways needed to maintain them become unreliable.

This issue mattered in the present case because the school website appeared to rely on WordPress-related infrastructure. In a setting such as Iran, where access to foreign services can become unstable during periods of protest, disruption, or war-related escalation, such dependence can create a practical asymmetry: the site may still be visible to users while becoming significantly harder for its maintainers to manage. The broader point is not that this one school site was itself globally consequential. It is that websites exist inside political and infrastructural environments that can change abruptly. Conflict, filtering, platform dependence, and sudden reforms can all affect not only whether a site is reachable, but whether it can continue to be actively maintained.

\subsection{Preservation infrastructure under pressure}

The project depended on the Internet Archive, yet the Internet Archive itself faces pressure. For many journalists, researchers, and investigators, the Wayback Machine is not a marginal tool but a practical part of modern reporting. It allows them to recover pages that have been deleted, compare how websites changed over time, verify what was publicly visible at an earlier moment, and preserve references that might otherwise disappear. Much of this use remains invisible to ordinary readers, who may not realize how often archival tools support verification and accountability behind the scenes.

Major publishers have begun restricting the Wayback Machine's crawlers, often out of concern that archived content may be used for AI training without permission \cite{niemanlab-blocks,eff-blocks,wired-blocks}. Even if one grants those concerns, the archival consequences are serious. Archiving and journalism have a  symbiotic relationship: archives preserve the historical record on which later reporting depends, while journalism demonstrates why preserving that record matters. When archival capture becomes harder, journalists lose part of the evidentiary infrastructure that helps them investigate, verify, and contextualize public claims.

As Warren Buffett once remarked, \enquote{The smarter the journalists are, the better off society is} \cite{buffett-journalists}. If archived reporting becomes harder to preserve and verify, society loses part of the public record on which informed judgment depends.

\section{Discussion}
\subsection{The web is structurally ephemeral}
This case study supports a broader claim: the ephemerality of the web is not a rare accident but a structural condition. Websites depend on maintainers, institutions, hosting providers, domains, political environments, and platform policies. Any of these can fail or change.

This means that the familiar intuition---if it is online, it will stay online---is wrong. Digital presence is not permanence. It is contingency.

\subsection{Archiving should become proactive}

If the web is structurally ephemeral, archiving cannot remain a niche activity performed only by a few specialists or hobbyists at the margins. But the solution is not indiscriminate or endlessly repeated archiving. Rather, archiving should become proactive: websites should be preserved at the moment they are meaningfully updated, redesigned, or overwritten.

My own project relied on repeated archival runs because no built-in preservation mechanism existed. In that sense, it was a workaround rather than an ideal model. Indeed, many links in the project were saved dozens of times, with some being archived more than fifty times. This repetition was not the result of an intrinsically desirable preservation strategy, but a symptom of a shortsighted archival ecosystem in which preservation is not integrated into ordinary website maintenance. In the absence of proactive mechanisms, outsiders are left repeatedly saving the same site in order to reduce the risk of loss.

That approach carries costs. It consumes time, bandwidth, and computing resources, and it still does not solve the underlying problem elegantly. A stronger model would be one in which preservation is tied directly to meaningful change. When institutions update public-facing pages, replace photographs, change domains, or redesign their sites, the previous state should be archived automatically as part of that process. In such a model, websites would not need to be saved over and over simply because no one could trust that important transitions were being preserved when they occurred.

By proactive archiving, I mean archiving tied to change. A school website should preserve its earlier version when a redesign occurs. A research group should preserve a project page when it is substantially revised. A content management system such as WordPress could, in principle, offer an option to archive a page whenever it is meaningfully updated. The guiding idea is simple: archiving should be integrated into ordinary website maintenance, so that historical states are preserved when they are replaced rather than repeatedly recovered afterward. If such a model were adopted widely, it would both reduce wasteful redundancy and make the web a little less ephemeral.

\subsection{Preservation as intergenerational responsibility}

\begin{displayquote}
\enquote{Life can only be understood backwards; but it must be lived forwards.}\\
--- Søren Kierkegaard
\end{displayquote}

One of the most meaningful parts of this project was its intergenerational motive. The goal was not simply to save data for myself, but to make sure that future students could see what existed before them. This is why archiving matters. It allows communities to retain continuity with their own past.

Digital memory is often discussed in abstract terms, but in practice it is deeply local. It can consist of something as specific as photographs of an older school building, an earlier homepage, or evidence of how an institution once presented itself. When such materials disappear, a community loses more than files; it loses part of its self-understanding.

The project also made me more aware of how brief our time is as caretakers of institutional memory. Individual lives and institutional roles are temporary, yet the digital traces we inherit and pass on can shape how later generations understand what came before them. For that reason, archiving is not only a technical practice but also an intergenerational one: it helps ensure that what is historically meaningful does not vanish simply because those who remember it directly are no longer present.

\section{Limitations}
This paper presents a single case study and a lightweight prototype rather than a comprehensive preservation framework.

First, the implementation focuses on one domain and uses site-specific link-discovery logic. It was designed for a particular preservation goal rather than as a general-purpose web crawler.

Second, the system submits URLs for archival capture but does not fully verify the completeness or rendering quality of each resulting archived page.

Third, the discussion of political and infrastructural constraints is contextual rather than a formal comparative network study.

Fourth, this paper does not attempt to resolve the legal and ethical questions surrounding archival reuse, especially in cases involving copyrighted, sensitive, or accidentally exposed material.

These limitations do not negate the value of the case. Rather, they help define the scope of the contribution: a practical demonstration and conceptual argument for proactive archiving grounded in one real deployment.

\section{Conclusion}
I began this project because my school changed its website, its URL, its principal, and its building within a short span of time, and because I did not want future generations to lose access to what had existed before. From that case emerged the broader argument of this paper: archiving should become proactive and integrated into ordinary website maintenance. My own project relied on repeated archival runs because no such mechanism was built in. In that sense, the project was less an ideal model than a symptom of a shortsighted preservation ecosystem: in the absence of proactive archiving, individuals at the margins may be forced to repeatedly save the same site in order to reduce the risk of historical loss.

The case study shows that preservation is technically feasible with modest tools, but also that ad hoc preservation remains vulnerable to platform limits, maintenance gaps, and political conditions. The larger lesson is not merely that one school website deserved to be saved. It is that if digital societies want to preserve institutional memory and public history without leaving preservation to chance, then archiving must become a proactive part of maintaining websites.

As the paper has suggested, communities live forward but understand themselves partly by looking backward. As more of human life shifts into digital form, that backward-looking understanding increasingly depends on digital archives. If we fail to preserve these materials now, we do not merely lose files in the present; we narrow the ability of future generations to understand who they were, what their institutions once were, and how their histories took shape.

\section*{Acknowledgments}
I am grateful to my friend Altay, whose conversation with me in December 2025 helped crystallize the intergenerational motivation behind this project. I am also grateful to the open-source communities and projects that shaped my thinking at Rensselaer Polytechnic Institute, especially RCOS, QuACS, and the broader examples of systems built to survive beyond their original student maintainers. Finally, I remain grateful to the school and community in Iran that gave rise to the memories this project sought to preserve. Even from far away, and even through periods of disruption, that connection remains. I finished this paper on 17 May 2026, three years to the day after the last day I attended the school. For those living through uncertainty now: everything will be alright in the end; and if it is not alright, it is not the end.

\end{document}